\documentclass{easychair}
\usepackage{hypdoc}

\usepackage{acronym}
\usepackage[style=numeric, maxnames=2]{biblatex}
\usepackage{doc}
\usepackage[inline,shortlabels]{enumitem}
\usepackage{todonotes}
\usepackage{booktabs}
\usepackage{uppaal}
\usepackage{commands}
\usepackage{comment}

\usepackage{tikz}
\usetikzlibrary{positioning}
\usepackage{acronym}

\acrodef{milt}[MITL]{Metric Interval Temporal Logic}
\acrodef{smc}[SMC]{Statistical Model Checking}
\acrodef{dsl}[DSL]{Domain-Specific Language}
\acrodefplural{dsls}[DSLs]{Domain-Specific Languages}
\acrodef{dfa}[DFA]{Deterministic Finite-state Automaton}
\acrodefplural{dfa}[DFA]{Deterministic Finite-state Automata}
\acrodef{sha}[SHA]{Stochastic Hybrid Automaton}
\acrodefplural{sha}[SHA]{Stochastic Hybrid Automata}
\acrodef{tctl}[TCTL]{Timed Computation Tree Logic}
\acrodef{poi}[PoI]{Point of Interest}
\acrodefplural{poi}[PoIs]{Points of Interest}
\acrodef{ode}[ODE]{Ordinary Differential Equation}
\acrodefplural{odes}[ODEs]{Ordinary Differential Equations}
\acrodef{ltl}[LTL]{Linear Temporal Logic}
\acrodef{hri}[HRI]{Human-Robot Interaction}
\acrodef{mas}[MAS]{Multi-Agent System}
\acrodefplural{mas}[MASs]{Multi-Agent Systems}
\acrodef{iot}[IoT]{Internet of Things}
\acrodef{liras}[LIrAs]{Language for Interactive Agents}
\acrodef{xml}[XML]{Extensible Markup Language}
\acrodef{ebnf}[EBNF]{Extended Backus–Naur Form}
\acrodef{bnf}[BNF]{Backus–Naur Form}

\acrodef{uml}[UML]{Unified Modeling Language}
\acrodef{mde}[MDE]{Model-Driven Engineering}
\acrodef{soc}[SoC]{State of Charge}
\acrodef{rq}[RQ]{Research Question}
\acrodef{llm}[LLM]{Large Language Model}
\acrodef{gt}[GT]{Ground Truth}
\acrodef{nl}[NL]{Natural Language}
\acrodef{nlp}[NLP]{Natural Language Processing}
\acrodef{ai}[AI]{Artificial Intelligence}
\acrodef{ctl}[CTL]{Computation Tree Logic}
\acrodef{ltl}[LTL]{Linear Temporal Logic}
\acrodef{tl}[TL]{Temporal Logic}
\acrodef{mitl}[MITL]{Metric Interval Temporal Logic}

\title{Towards an Agentic LLM-based Approach to Requirement Formalization from Unstructured Specifications}

\author{
Alberto Tagliaferro\inst{1}
\and
Bruno Guindani\inst{1}
\and
Livia Lestingi\inst{1}
\and
Matteo Rossi\inst{1}
}
\institute{
  Politecnico di Milano,
  Milan, Italy\\
  \email{name.surname@polimi.it}
 }
 
\authorrunning{Tagliaferro, Guindani, Lestingi, and Rossi}

\titlerunning{}

\begin{document}

\maketitle

\begin{abstract}

Early-stage specifications of safety-critical systems are typically expressed in natural language, making it difficult to derive formal properties suitable for verification and needed to guarantee safety.
While recent \ac{llm}–based approaches can generate formal artifacts from text, they mainly focus on syntactic correctness and do not ensure semantic alignment between informal requirements and formally verifiable properties.
We propose an agentic methodology that automatically extracts verification-ready properties from unstructured specifications. The modular pipeline combines requirement extraction, compatibility filtering with respect to a target formalism, and translation into formal properties. 
Experimental results across three scenarios show that the pipeline generates syntactically and semantically aligned formal properties with a $77.8\%$ accuracy.
By explicitly accounting for modeling and verification constraints, the approach is a paving step towards exploiting \ac{ai} to bridge the gap between informal descriptions and semantically meaningful formal verification.

\subsubsection*{Keywords}
Formal Property Elicitation, Requirement Formalization, AI for Requirements Engineering, AI for Formal Methods, Large Language Models, Agentic Workflow
\end{abstract}

\acresetall

\section{Introduction}
\label{sec:intro}

When engineering software-intensive safety-critical systems, providing sound guarantees that the delivered product satisfies set requirements is of pivotal importance.
For example, for service robots requiring close interaction with humans to complete their tasks, it is crucial to ensure that their behavior complies with both safety and societal norms.
Model-driven approaches in combination with formal verification techniques are traditionally adopted to guarantee that the final software-controlled system will satisfy the desired properties.
To this end, a formal specification of the system under analysis, amenable to verification, and of the properties to be satisfied must be available. 
However, at an early stage, the software development process is usually carried out by means of \emph{informal} specifications, such as natural language descriptions of the \ac{hri} scenarios that the robotic agent must be able to tackle, and eliciting verifiable requirements from free text remains an open challenge.

The existing body of work on requirement elicitation widely explores the issue of extrapolating meaningful requirements from unstructured text with an automated process \cite{meth2013state,lim2021data}. 
On the other hand, the goal of automatically \emph{formalizing} requirements out of natural language descriptions has only recently been pursued with the advent of generative \ac{ai} and, specifically, \acp{llm}.  
However, recent work mostly focuses on approaches that ensure the \emph{syntactic} correctness of the \ac{ai}-generated artifacts. 
Syntactic correctness can be checked automatically by validating models against a grammar, compiling them, or verifying well-formedness constraints defined by modeling guidelines. 
While these checks ensure structural consistency, they do not guarantee that the generated requirements also \emph{semantically} align with the intended system's behavior.
Verifying semantic correctness is substantially more difficult: properties must reflect the intent expressed in natural language specifications, be expressible in a formal language, and be verifiable on the resulting model.

This paper addresses this challenge by proposing a methodology that automatically extracts a set of formal properties from an informal specification, with the goal of operationalizing requirements that are not only syntactically but also semantically correct with respect to the initial specification.
The approach explicitly accounts for the characteristics and constraints of the target model, ensuring that the generated properties are meaningful and verifiable.
Thanks to the modular structure of the methodology, it is possible to trace and inspect each stage of the property elicitation process.

The presented methodology proposes an agentic approach based on \acp{llm} to elicit verification-ready properties from unstructured textual specifications. 
The approach is \emph{agentic} as it coordinates multiple \acp{llm} with specialized roles and integrates verification tools within the pipeline, allowing candidate properties to be extracted, assessed, and translated while being validated against the constraints of the target verification framework.
Specifically, the pipeline envisages the following stages:
\begin{enumerate*}[label=(\roman*)]
    \item an \ac{llm} extracts candidate natural language requirements from informal specifications,
    \item candidate requirements are then filtered by a judge \ac{llm} to exclude the ones corresponding to properties that are not compatible with or cannot be expressed in the target formal language, 
    \item eligible candidates are translated into properties expressed in the target formalism, and,
    \item finally, the syntactic correctness of the generated properties is programmatically checked through the target verification tool while a second judge \ac{llm} evaluates the semantic alignment with the original specification.
\end{enumerate*}

While the approach is domain-agnostic, our preliminary evaluation focuses on informal specifications in the \ac{hri} domain.
Each specification is translated into a \ac{liras} specification, a \ac{dsl} designed for describing multi-agent interaction patterns \cite{tagliaferro2024towards,tagliaferro2024verification}, which automatically generates a \ac{sha} model compatible with the \uppaal tool.\footnote{\href{https://www.uppaal.org}{www.uppaal.org}} Consequently, the target formalism for this work is the query language provided by \uppaal in support of \ac{smc}, but as mentioned, the proposed pipeline can be applied to different domains, formalisms, and verification tools.
Experiments are performed with the \emph{gemini-3.1-pro-preview} model.\footnote{The code and artifacts for this project are publicly available at \href{https://doi.org/10.5281/zenodo.19023550}{doi.org/10.5281/zenodo.19023550}.}
Results show that:
(i) in the extraction phase, the model achieves an $81.8\%$ combined rate of exact and partial semantic matches for the generated requirements; 
(ii) in the verifiability classification stage, it has an
overall accuracy of $88.7\%$ and a recall of $94.2\%$, successfully filtering out unverifiable constraints; and 
(iii) in the final formal translation step, $95.8\%$ of the generated queries are syntactically correct, with a $77.8\%$ effective semantic translation accuracy when accounting for valid logical equivalences and alternative formulations.

This paper is structured as follows:
\sref{related} surveys related work,
\sref{bg} outlines preliminary concepts,
\sref{method} describes the proposed methodology,
\sref{evaluation} presents experimental results, and
\sref{concl} concludes and presents future research directions.

\section{Related work}
\label{sec:related}

This section surveys related work on requirement formalization from informal specifications and the use of \acp{llm} in requirements engineering, positioning our work with respect to the state-of-the-art.

\sloppypar{\noindent\textbf{Requirement Formalization from Informal Specifications.}}
\citeauth{beg2025leveraging} survey \mbox{(semi-)automated} approaches for generating formal specifications from \ac{nl} requirements, including \ac{nl}, ontology-based domain modeling, and \acp{llm}, and discuss key challenges identified across the literature.
These include semantic ambiguity, the absence of ground-truth datasets, limited tool interoperability, traceability across artifacts, and concerns regarding explainability and user trust. 

Recent work increasingly leverages \acp{llm} for direct translation into temporal logics.
For instance, \citeauth{li2025extracting} generate \ac{tl} specifications from software documentation using both end-to-end and two-step pipelines, reporting issues of specification oversimplification and fabrication.
\citeauth{cosler2023nl2spec} introduce nl2spec, an open-source framework that translates unstructured \ac{nl} requirements into \ac{tl} statements via \acp{llm}, while \citeauth{zhao2024nl2ctl} focus specifically on automated conversion into \ac{ctl}.

Several approaches introduce intermediate representations to structure the translation process.
In \citeauth{ma2025bridging}, \acp{llm} map informal requirements to a JSON-like language, OnionL, which is subsequently translated into \ac{ltl} through rule-based techniques.
Similarly, \citeauth{guidotti2024translating} convert \ac{nl} requirements into Property Specification Patterns (PSP), an intermediate formalism enabling translation into multiple target logics.

Beyond temporal logic generation, \citeauth{wen2024enchanting} and \citeauth{faria2026automatic} employ \ac{llm}-based techniques to synthesize pre- and post-condition annotations in various programming languages.
Related research also explores the generation of formal theorem statements and proofs, notably the autoformalization work of \citeauth{wu2022autoformalization}.
However, while these approaches address the correctness or plausibility of generated formal artifacts, they do not explicitly evaluate their consistency with the originating informal specification.

Compared to rule-based requirement extraction or \ac{nlp} pipelines designed to derive verifiable properties, \ac{llm}-based approaches offer greater flexibility and portability.
Traditional methods often require substantial customization for each domain, modeling language, or verification framework.
In contrast, the proposed methodology can be adapted with minimal effort: changes typically involve revising prompt instructions and textual examples provided to the \ac{llm}, rather than redesigning extraction rules or retraining specialized models.
This makes the pipeline a practical and extensible solution for bridging the gap between informal system descriptions and formal, semantically meaningful verification properties.

\sloppypar{\noindent\textbf{\ac{llm}-Assisted Requirements Engineering.}}
The usage of \acp{llm} to assist software engineering is a relevant subject in the state of the art as well as the industry \cite{jahic2024state}.
Consistency checking has long been a central topic in requirements engineering (see, \eg{} \citeauth{yan2015formal}), but \acp{llm} have recently expanded the methodological landscape.
The survey by \citeauth{zadenoori2025large} shows that ``most of the studies focus on using \acp{llm} for requirements elicitation and validation, rather than defect detection and classification'', where \emph{validation} denotes assessing whether ``requirements accurately reflect stakeholder intent''.
However, validation in this sense does not typically include verifying that derived requirements remain coherent with a given textual system specification.

Work on ``defect detection and classification'' (as categorized by \citeauth{zadenoori2025large}) likewise omits this perspective.
These studies focus primarily on inconsistencies \emph{between} requirements, logical flaws within individual requirements, or violations of writing best practices.
For example, \citeauth{fazelnia2024lessons} address
\begin{enumerate*}[label=\textit{\roman*)}]
\item classification into categories such as security, performance, and portability,
\item detection of defects such as ambiguity or non-compliance with requirements-writing guidelines, and
\item identification of inter-requirement conflicts.
\end{enumerate*}
Similarly, \citeauth{mahbub2024can} examine ambiguity, inter-requirement inconsistencies, and incompleteness in large industrial requirement sets, while \citeauth{luitel2024improving} use \acp{llm} as external knowledge sources to detect incompleteness in \ac{nl} requirements.
\citeauth{bertram2023leveraging} focus on pattern classification and translation into Structured English with stricter grammatical constraints.
Finally, \citeauth{gartner2024automated} and \citeauth{chen2025llm} investigate contradiction detection \emph{between} requirements, the former by combining formal reasoning decision tree-based frameworks with \acp{llm} and the latter by generating imperative SMT-based code checkers.

While existing approaches emphasize syntactic correctness, formal well-formedness, or successful translation into verification artifacts, they provide limited support for assessing whether the generated formal properties faithfully capture the intent of the original informal specification.
Explicitly evaluating this semantic consistency is the central objective of our work.

\section{Preliminaries}
\label{sec:bg}

In the following, we outline preliminary concepts underlying our work, \ie \acp{llm}, few-shot learning, and \uppaal's \ac{smc} query language. 

\sloppypar{\textbf{\acp{llm} and Few-shot Learning.}}
An \ac{llm} is an autoregressive and probabilistic generative model for text trained on massive amounts of data. 
\acp{llm} are built upon the \emph{Transformer} architecture that processes sequences of text tokens through the \emph{self-attention} mechanism~\cite{DBLP:conf/nips/VaswaniSPUJGKP17}.
As autoregressive models, \acp{llm} work by predicting, one token at a time, the most likely new token based on all the context at hand and all the previously generated tokens.

When \emph{prompted} with some text, \acp{llm} generate a completion (\ie the response) for the given prompt.
To explicitly steer this generation, a \emph{system prompt} is often employed to define an overarching persona, specify operational boundaries, and dictate the general rules the \ac{llm} must follow throughout the interaction.
Furthermore, several approaches have been developed to structure the prompt (also referred to as prompting \emph{strategies}): \emph{few-shot learning} (sometimes called in-context learning~\cite{DBLP:conf/nips/BrownMRSKDNSSAA20}) is among the most commonly adopted strategies.
Few-shot learning envisages incorporating into the prompt examples of input-output pairs (\ie the \emph{shots}) together with the task instructions. 
This helps the \ac{llm} identify patterns in the text and clarifies the expected input/output formats~\cite{DBLP:conf/nips/BrownMRSKDNSSAA20}.

\sloppypar{\textbf{\uppaal \ac{smc}.}}
Stochastic formal models are amenable to \ac{smc} \cite{agha2018survey}, which can be performed through \uppaal \cite{david2015uppaal}.
\ac{smc} applies statistical techniques to a set of \emph{runs} of a formal model $M$ with stochastic features to estimate the \emph{probability} of the desired property holding.
\uppaal specifically computes the value of expression $\mathbb{P}_M(\psi)$ to estimate the probability of property $\psi$ holding for model $M$ \cite{david2015uppaal}. 
Property $\psi$ is expressed in \ac{mitl} \cite{alur1996benefits}, a linear temporal logic interpreted over timed state sequences that allows users to capture real-time constraints.
In particular, \ac{mitl} includes operator $\lozenge_{[a, b]}$ (with $a < b$), where formula $\lozenge_{[a, b]} \phi$ means that property $\phi$ holds at a point in the future (i.e., ``eventually'') that is no less than $a$ and no more than $b$ time units from the current one.
Specifically, $\psi$ in expression $\mathbb{P}_M(\psi)$ is
of the form $\lozenge_{\leq\tau}\ \mathsf{ap}$, where
$\lozenge_{\leq\tau}$ is an abbreviation for $\lozenge_{[0,\tau]}$
and ${\mathsf{ap} \in \mathrm{AP}}$ is an atomic proposition.
Formula $\lozenge_{\leq\tau}\ \mathsf{ap}$ is true in the first instant of a run of model $M$ if $\mathsf{ap}$ holds within $\tau$ time units from time instant $0$.
In the \uppaal \ac{smc} query language, $\mathbb{P}_M(\lozenge_{\leq\tau}\ \mathsf{ap})$ is expressed as \texttt{P[<=$\tau$](<> \textsf{ap})}, where \textsf{ap} is either an atomic proposition or a conjunction/disjunction thereof.

\section{Methodology}
\label{sec:method}

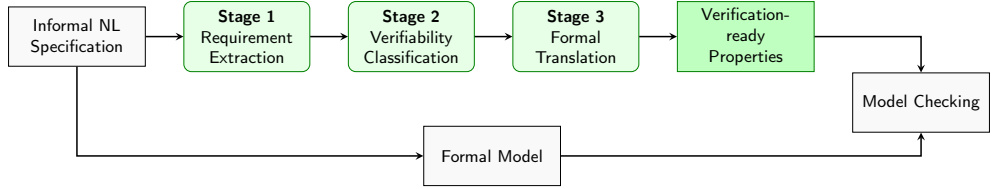
\begin{figure}[t]
    \centering
    \resizebox{0.9\columnwidth}{!}{%
    \begin{tikzpicture}[
        font=\sffamily\small,
        >=stealth,
        stage/.style={rectangle, draw=green!50!gray!200, rounded corners, fill=green!10, text width=2.1cm, align=center, minimum height=1.3cm},
        data/.style={rectangle, draw=black, fill=gray!5, text width=2.3cm, align=center, minimum height=1.1cm},
        arrow/.style={->, thick},
        data2/.style={rectangle, draw=green!50!gray!200, fill=green!25, text width=2.3cm, align=center, minimum height=1.1cm},
        arrow/.style={->, thick}
    ]
        \node[data] (nl) {Informal NL\\Specification};
        
        \node[stage, right=0.7cm of nl] (s1) {\textbf{Stage 1}\\Requirement\\Extraction};
        
        \node[stage, right=0.7cm of s1] (s2) {\textbf{Stage 2}\\Verifiability\\Classification};
        
        \node[stage, right=0.7cm of s2] (s3) {\textbf{Stage 3}\\Formal\\Translation};
        
        \node[data2, right=0.7 cm of s3] (queries) 
        {Verification-ready\\Properties};

        \node[data, below=1cm of queries, xshift=-4.7cm] (fm) {Formal Model};
        
        \node[data, below right=.01cm and .7cm of queries] (ver) {Model Checking};

        \draw[arrow] (nl) -- (s1);
        \draw[arrow] (s1) -- (s2);
        \draw[arrow] (s2) -- (s3);
        \draw[arrow] (s3) -- (queries);
        
        \draw[arrow] (nl.south) |- (fm.west);
        \draw[arrow] (queries.east) -| (ver.north);
        \draw[arrow] (fm.east) -| (ver.south);

    \end{tikzpicture}%
    }
    \caption{Proposed agentic pipeline. In green, the part explored by this work.}
    \label{fig:pipeline_architecture}
\end{figure}

This section details the proposed agentic pipeline for bridging the gap between informal \ac{nl} specifications and formal semantically aligned properties. 
The approach is structured into three distinct stages to ensure traceability, mitigate the overall complexity, and improve the performance of \acp{llm}.
The pipeline is illustrated in Figure~\ref{fig:pipeline_architecture}, where \textbf{Stage 1}, \textbf{Stage 2} and \textbf{Stage 3}, corresponds to three single agent \ac{llm}-based step.

\sloppypar{\textbf{Stage 1: Requirements Extraction.}}
The first stage transforms unstructured informal specifications into a structured set of atomic natural-language requirements, each one associated with an identifier, isolating functional and safety constraints embedded in the system description, while using the standard requirements syntax (\eg "The system must...", "The human agent has to...", "The robot shall...").

An \ac{llm} is used to identify relevant entities, system dynamics, and operational constraints described in the specification.
The extracted requirements are normalized and structured into a machine-readable representation (\ie JSON format), enabling their systematic processing in the subsequent stages of the pipeline.

\sloppypar{\textbf{Stage 2: Verifiability Classification.}}
The second stage filters requirements that cannot be expressed within the semantic boundaries of the target formal model.
Indeed, not all elicited requirements can be formally verified within the constraints of the considered model.
To perform this classification, we use an \ac{llm}, as its adaptability across different domains facilitates reuse; moreover, due to its nature, it is capable of performing classification tasks based on semantic evaluation.
Thanks to this second classification stage, in the first one, the \ac{llm} focuses solely on generating a complete set of requirements without bias that can be derived by directly considering the limitations of the specified formal model.

In this stage, each requirement is analyzed with respect to the modelling assumptions and observable variables of the system.
Requirements referring to unobservable properties, qualitative design intentions, or model-external phenomena are discarded. The remaining requirements constitute the subset that can be formally analyzed.

\sloppypar{\textbf{Stage 3: Formal Translation.}}
The final stage translates the filtered requirements into formal verification queries compatible with the target formal model.
The objective of this stage is to bridge the gap between natural language specifications, which are inherently ambiguous, and formal verification artifacts.

Thanks to the ability of \acp{llm} to generate text according to a precise grammar (provided in BNF format in the system prompt), each requirement is translated into a corresponding temporal property that can be evaluated on the system model.

The resulting queries represent formalized constraints that can be directly analyzed through model checking, providing a formalization of the otherwise ambiguous semantics inherent in natural language requirements.

\section{Experimental Evaluation}
\label{sec:evaluation}

In this section, we evaluate each stage of the proposed pipeline to assess its effectiveness in translating informal \ac{nl} specifications into formal verification queries. For this evaluation, we used the \ac{liras} \ac{dsl} to generate the formal model corresponding to the informal specification. In this setting, the target formalism is \acp{sha}, and verification is performed using \uppaal \ac{smc} queries.
Our evaluation is designed to answer the following \acp{rq}:

\begin{enumerate}[label=\textbf{RQ\arabic*.}, itemsep=2pt, parsep=0pt]
    \item \textbf{Requirements Extraction:} How accurately can the \ac{llm} extract requirements from informal \ac{nl} specifications?
    \item \textbf{Verifiability Classification:} How effectively can the \ac{llm} classify the formal verifiability of requirements with respect to strict model boundaries?
    \item \textbf{Formal Translation:} To what extent can the \ac{llm} translate verifiable requirements into syntactically correct and semantically equivalent \uppaal \ac{smc} queries?
\end{enumerate}

\subsection{Experimental Setup}
\label{subsec:setup}

The evaluation is conducted on a dataset comprising three distinct cyber-physical system scenarios (SC): \emph{SC1: Coffee\_Delivery}, \emph{SC2: User\_Guided\_Transport}, and \emph{SC3: Factory\_Pipeline}, inspired by the literature \cite{menghi2019specification,bajones2019results,IFR}.

These specifications were translated into the \ac{liras} \ac{dsl}, from which the corresponding \acp{sha} models were automatically generated. Table~\ref{tab:experimental_setup} summarizes the dataset size and associated \ac{gt} information. For example, the first scenario takes as input one agent, three locations, and one resource. From the informal specification, we manually constructed the \ac{gt}, obtaining $19$ \ac{nl} requirements. Of these, $12$ are suitable for formal verification, producing $15$ \ac{smc} queries (as a requirement may correspond to multiple queries). Non-verifiable requirements arise mainly from model abstractions or technical limitations of \ac{smc}.

\begin{table}[t]
    \centering
    \caption{Overview of the dataset size. \emph{Ag.} denotes the number of agents defined in the LIrAs code, \emph{Loc.} denotes the number of locations, and \emph{Res.} denotes the number of resources. \emph{Req.} stands for requirements, and \emph{Ver.} stands for verifiable.}
    \label{tab:experimental_setup}
    \begin{tabular}{l c c c c c c}
        \toprule
        \textbf{Scenario} & \textbf{Ag.} & \textbf{Loc.} & \textbf{Res.} & \textbf{GT Req.} & \textbf{GT Ver. Req.} & \textbf{GT Queries} \\
        \midrule
        \textbf{SC1}   & 1 & 3 & 1 & 19 & 12 & 15 \\
        \textbf{SC2}   & 2 & 2 & 1 & 26 & 22 & 23 \\
        \textbf{SC3}   & 2 & 6 & 3 & 26 & 18 & 34 \\
        \midrule
        \textbf{Total} & \textbf{-} & \textbf{-} & \textbf{-} & \textbf{71} & \textbf{52} & \textbf{72}  \\
        \bottomrule
    \end{tabular}
\end{table}

To ensure a rigorous evaluation, each pipeline stage was executed and analyzed independently using the \ac{gt} as the benchmark. This decoupled setup measures the performance of each component while preventing cascading errors.

\paragraph{RQ1: Requirements Extraction.}
In this first stage, a system prompt instructs the \ac{llm} to extract atomic requirements from 
informal specifications. The prompt directs the model to focus on physical dynamics, safety aspects, and agent roles, while constraining the output to a strict \emph{JSON} schema to ensure seamless integration with downstream pipeline steps.

We employ a 2-shot prompting strategy in which, for each scenario, the \acp{gt} of the other two scenarios are used as examples.
Specifically, the user prompt provides the informal specification, while the assistant prompt supplies the corresponding hand-crafted requirements. This setup guides the model to correctly structure the generated requirements and format them according to the expected \emph{JSON} output.

To evaluate the quality of this step, we compare each generated set of requirements against the corresponding ground truth. Since the two lists are difficult to compare directly---due to the inherent variability of informal requirements---we adopt an \ac{llm}-as-a-judge approach. The judge model is instructed through a system prompt to perform a semantic evaluation, allowing many-to-many matching between generated and ground-truth requirements.
It applies the following three classification rules:

\begin{itemize}[itemsep=2pt, parsep=0pt]
    \item[-] Match: Same intent, scope, and constraints; differences are purely stylistic.
    \item[-] Partial: The intent is similar, but the scope differs (e.g., more specific or more abstract).
    \item[-] NoMatch: No \ac{gt} requirement captures the semantic intent of the generated requirement.
\end{itemize}

Additionally, the judge provides a confidence score (between $0.0$ and $1.0$), a short justification for each match, and identifies any \ac{gt} requirements that were missed.

The judge receives the informal specification (for context) and the two requirement lists (generated and ground truth), operating in a 0-shot setting.

\paragraph{RQ2: Verifiability Classification.}
In the second stage, the pipeline acts as a formal gatekeeper. 
We instruct the \ac{llm} to assume the role of an \emph{expert systems engineer} specializing in \emph{formal methods} and \emph{\ac{smc}}. 
The task is to evaluate the requirements and determine whether each one is formally verifiable within the strict boundaries of the specific \acp{sha} model. To ensure independence from the previous step, we use the \ac{gt} requirements as input for this stage.

The \ac{llm} is provided with the informal specification and the list of requirements to be evaluated. To reduce hallucinations, the system prompt explicitly defines the semantic boundaries of the \ac{liras} model, including assumptions such as the system being memoryless, the absence of physical damage detection, and the assumption that stationary machines have infinite resources and zero processing delay.

The \ac{llm} must strictly apply a set of structured classification rules to filter out unverifiable concepts, such as unobservable variables or qualitative design choices. The expected output is a strict \emph{JSON} array that classifies each requirement as either ``Yes'' (verifiable) or ``No'' (not verifiable), together with a concise justification, helping the \ac{llm} in the reasoning process.
To evaluate this step, we compare the \ac{llm}'s ``Yes/No'' predictions with the manually defined \ac{gt} verifiability labels.

\paragraph{RQ3: Formal Translation.}
The final step translates verifiable requirements into syntactically correct \uppaal \ac{smc} queries that can be verified on the model generated from the \ac{liras} \ac{dsl} for the associated specification. The system prompt enforces strict constraints, requiring the generated queries to conform to a compact BNF grammar for \ac{smc}. Furthermore, the model is guided by explicit mapping rules that bridge the \ac{liras} BNF grammar and the corresponding \uppaal model.
To evaluate the generated queries, we adopt a rigorous three-level strategy:

\begin{enumerate}[itemsep=2pt, parsep=0pt]
    
    \item \emph{Syntax Check:} First, we evaluate the syntactic validity of the queries by passing them through the \uppaal \texttt{verifyta} compiler. Any query that fails to compile is marked as a failure, preventing the propagation of invalid syntax.
    
    \item \emph{Exact Match:} For queries that correctly compile, we check for perfect textual equivalence between the generated queries and the \ac{gt} queries.
    
    \item \emph{Semantic Evaluation:} For queries that successfully compile but differ textually from the \ac{gt}, we employ an \ac{llm}-as-a-judge to assess semantic equivalence.
    The judge is instructed with specific equivalence rules, such as logical commutativity, spatial symmetry in distance functions, and tolerance for minor variations in \ac{smc} simulation time bounds. Conversely, it is explicitly instructed to penalize critical mismatches, such as confusing reachability with safety invariants.
    While some mathematical equivalences could in principle be verified through stricter formal rewriting rules, the \ac{llm} is particularly useful in identifying higher-level logical correspondences between queries that arise from different modeling perspectives. For instance, verifying that an agent reaches a target location can be expressed either by directly checking the agent's coordinates or by verifying the completion of the corresponding navigation action.
    The \ac{llm}-based judge allows us to capture such semantically equivalent formulations that are difficult to encode through purely syntactic comparison rules. The judge outputs a Boolean match flag together with a concise justification, enabling the computation of a relaxed, yet formally grounded, semantic accuracy metric.
\end{enumerate}

All experimental evaluations were conducted using the \emph{gemini-3.1-pro-preview} model. To reduce the inherent randomness of the language model, the temperature parameter was set to $0.1$. Additionally, the model output was strictly constrained to the \emph{application/json} format to ensure automated parsing and compliance with the provided validation schemas.

\subsection{Results}

\paragraph{RQ1: Requirements Extraction.}
\label{subsubsec:rq1_results}
To answer RQ1, we evaluate how accurately the pipeline extracts structured atomic requirements from informal specifications by comparing the generated requirements against the manually constructed \ac{gt}.
As shown in Table~\ref{tab:rq1_extraction}, the model achieved an exact semantic match for $43.6\%$ of the generated requirements, a partial match for $38.2\%$, and no match for $18.2\%$.
The recall of the captured \ac{gt} requirements is $60.6\%$, with $28$ out of the total $71$ requirements not mapped to any generated requirement.


\begin{table}[t]
    \centering
    \caption{Distribution of requirement extraction accuracy across the evaluated scenarios.}
    \label{tab:rq1_extraction}
    \resizebox{\textwidth}{!}{
        \begin{tabular}{l c c c c c c}
            \toprule
            \textbf{Scenario} & \textbf{GT Req.} & \textbf{Gen. Req.} & \textbf{Match} & \textbf{Partial} & \textbf{NoMatch} & \textbf{Missed GT} \\
            \midrule
            \textbf{SC1} & 19 & 16 & 7 (43.8)\% & 7 (43.8)\%  & 2 (1.25)\% & 6 (31.6)\% \\
            \textbf{SC2} & 26 & 15 & 9 (60.0)\% & 3 (20.0)\%  & 3 (20.0)\% & 15 (57.7)\% \\
            \textbf{SC3} & 26 & 24 & 8 (33.3)\% & 11 (45.8)\% & 5 (20.8)\%& 7 (26.9\%) \\
            \midrule
            \textbf{Total} & \textbf{71} & \textbf{55} & \textbf{24 (43.6\%)} & \textbf{21 (38.2\%)} & \textbf{10 (18.2\%)} & \textbf{28 (39.4\%)} \\
        \end{tabular}
    }
    \vspace{1ex}
    \\
    \raggedright
\end{table}

Across the evaluated scenarios, the generated requirements consistently capture the core operational logic, including basic navigation and overall task sequences (\eg object transport, synchronization, and deployment). However, the qualitative analysis reveals two recurring failure modes in the extraction process.

First, the \ac{llm} frequently omits strict logical boundaries and formal prerequisites present in the \ac{gt}, such as conditions preventing infinite loops, guarantees of task completion, and mechanisms for failure detection. Notably, \emph{SC2} represents a negative outlier in this regard: the model completely missed fundamental state transitions and physical constraints (\eg battery charge and human fatigue) that were successfully captured in \emph{SC1} and \emph{SC3}.

Second, the model shows a persistent tendency to introduce unstated but plausible constraints derived from general domain knowledge rather than from the provided text. Examples include explicitly identifying employees in \emph{SC1} and enforcing safe following distances in \emph{SC2}. Furthermore, as observed in \emph{SC3}, the \ac{llm} struggles with maintaining the intended level of abstraction, often decomposing high-level transport operations into highly granular manipulation steps. This behavior increases the number of partial matches, as the generated requirements diverge in scope from the \ac{gt}.

\vspace{0.15cm}
\noindent
\setlength{\fboxsep}{5pt}
\setlength{\fboxrule}{.7pt}
\fcolorbox{gray!95}{gray!10}{%
    \parbox{0.93\columnwidth}{%
        \textbf{RQ1 summary.} 
        The \ac{llm} effectively extracts the core functional logic and navigation sequences from informal specifications. However, it struggles with formal rigor, often omitting strict logical constraints and occasionally hallucinating plausible but unmodeled domain behaviors. Overall, the combined rate of exact and partial matches reaches $81.8\%$ respect to the total number of generated requirements.
    }
}

\paragraph{RQ2: Verifiability Classification.}
\label{subsubsec:rq2_results}
For \textbf{RQ2}, we evaluate the ability of the pipeline to filter out unverifiable requirements based on the system's structural constraints. Table~\ref{tab:rq2_confusion} summarizes the classification performance. The model achieved an overall accuracy of $88.7\%$ (ranging from $84.2\%$ to $92.3\%$ across the evaluated scenarios), correctly classifying $63$ out of $71$ requirements. The precision is $90.7\%$, while the recall reaches $94.2\%$.

\begin{table}[t]
    \centering
    \caption{Confusion matrix for verifiability classification (Yes/No). TP = True Positive, FN = False Negative, FP = False Positive, TN = True Negative.}
    \label{tab:rq2_confusion}
    \begin{tabular}{lcc}
        \toprule
         & \textbf{Predicted Verifiable} & \textbf{Predicted Unverifiable} \\
        \midrule
        \textbf{Actual Verifiable} & $TP=49$ & $FN=3$ \\
        \textbf{Actual Unverifiable}& $FP=5$ & $TN=14$ \\
    \end{tabular}
\end{table}

These results highlight the importance and effectiveness of this intermediate gatekeeping step. The high recall indicates that the \ac{llm} rarely discards valid, formalizable constraints, thereby preserving the core functional requirements of the system.

Importantly, the model correctly identified and filtered out $14$ unverifiable requirements (true negatives). This pre-filtering mechanism is essential for the overall efficiency and robustness of the pipeline. By discarding qualitative goals, unobservable properties, or structurally incompatible requirements early in the process, the system reduces token consumption during the computationally expensive generation of \uppaal queries. More importantly, it prevents downstream failures by ensuring that the \uppaal \texttt{verifyta} compiler is not burdened with mathematically inexpressible queries, or unmodelled variables, that would inevitably trigger syntax or compilation errors.

Although the model missed some unverifiable requirements 
($5$ false positives), the high precision ($90.74\%$) keeps this leakage limited. Furthermore, the pipeline architecture is resilient to such minor misclassifications, since any erroneous inclusions at this stage will be detected during the strict syntactic and semantic validation performed in the final translation step.

\vspace{0.15cm}
\noindent
\setlength{\fboxsep}{5pt}
\setlength{\fboxrule}{.7pt}
\fcolorbox{gray!95}{gray!10}{%
    \parbox{0.93\columnwidth}{%
        \textbf{RQ2 summary.} The \ac{llm} classifies requirement verifiability with $88.7\%$ accuracy and retains more than $94\%$ of the requirements that correspond to valid formal constraints, with respect to our \ac{gt}.
    }
}

\paragraph{RQ3: Formal Query Translation.}
\label{subsubsec:rq3_results}
This question investigates the generation of formal \ac{smc} queries.
To evaluate this capability, we adopt a three-level evaluation methodology: (i) syntactic validation using the \uppaal verification engine, 
(ii) strict string matching against the \ac{gt} queries, and 
(iii) semantic equivalence evaluation using an \ac{llm}-as-a-judge for syntactically valid but textually divergent queries.
The results of these evaluations are reported in Table~\ref{tab:rq3_translation}.

Out of the $72$ generated queries, $69$ ($95.8\%$) successfully compiled in \uppaal, indicating a strong adherence to both the DSL structure and the \uppaal BNF grammar. However, only $25$ queries achieved an exact textual match with the \ac{gt}. After the semantic evaluation, the \ac{llm}-based judge assessed the remaining syntactically valid queries and identified $31$ additional queries as semantically correct. As a result, the overall relaxed accuracy increased to $77.8\%$.


\begin{table*}[t]
    \centering
    \caption{Evaluation of formal translation into \uppaal SMC queries.}
    \label{tab:rq3_translation}
    \resizebox{\textwidth}{!}{
        \begin{tabular}{l c c c c c}
            \toprule
            \textbf{Scenario} & \textbf{Queries} & \textbf{Compiled Syntax} & \textbf{Exact Match} & \textbf{LLM Judged Valid} & \textbf{Accuracy} \\
            \midrule
            \textbf{SC1} & 15 & 15 (100\%) & 8 (53.3\%) & 3 (42.9\%) & \textbf{11 (73.3\%)} \\
            \textbf{SC2} & 23 & 22 (95.7\%) & 10 (43.5\%) & 3 (25\%) & \textbf{13 (56.5\%)} \\
            \textbf{SC3} & 34 & 32 (94.1\%) & 7 (20.6\%) & 25 (100\%) & \textbf{32 (94.1\%)} \\
            \midrule
            \textbf{Total} & \textbf{72} & \textbf{69 (95.8\%)} & \textbf{25 (34.7\%)} & \textbf{31 (70.5\%)} & \textbf{56 (77.8\%)} \\
        \end{tabular}
    }
    \raggedright
\end{table*}

These results highlight a well-known limitation when evaluating code or formal query generation: strict string matching (exact match) is an inherently limited and overly restrictive metric. While only $34.7\%$ of the generated queries exactly match the \ac{gt}, this metric does not accurately reflect the model's logical capabilities, since multiple syntactically different expressions can represent the same mathematical semantics.

Our qualitative analysis of the rejected exact matches shows that the \ac{llm} frequently produced valid formal variations. For example, the model often applied logical commutativity (\eg evaluating $A \land B$ instead of $B \land A$), implication equivalences (\eg expressing $A \implies B$ as $\neg A \lor B$), or leveraged spatial symmetry in Cartesian distance functions (\eg computing the distance from the human to the robot rather than from the robot to the human).

While some of these equivalences could be captured through deterministic rewriting rules, the main challenge lies in identifying semantically equivalent queries that arise from different modeling abstractions. For instance, verifying that an agent completes an action can be expressed either by directly checking some physical parameters or by verifying the completion state of the corresponding action orchestrated by the system. Such formulations are logically equivalent at the specification level but may differ significantly in their syntactic structure.

The semantic evaluation step is therefore critical. Using an \ac{llm}-as-a-judge guided by formal equivalence rules, we identified $31$ additional queries functionally equivalent to the \ac{gt}. Its impact is particularly evident in \emph{SC3}: while exact matching yields only $20.6\%$ accuracy, the model often produces valid alternative formulations, such as checking a sequence orchestrator's completion state instead of raw spatial coordinates. Accounting for these equivalents validates $25$ additional queries, raising \emph{SC3}'s semantic accuracy to $94.1\%$ and overall accuracy to $77.8\%$.

\vspace{0.15cm}
\noindent
\setlength{\fboxsep}{5pt}
\setlength{\fboxrule}{.7pt}
\fcolorbox{gray!95}{gray!10}{%
    \parbox{0.93\columnwidth}{%
        \textbf{RQ3 summary.} $95.8\%$ of the \ac{llm}-generated verification-ready queries are syntactically correct.
        Furthermore, semantic evaluation shows that strict string matching significantly underestimates performance: accounting for logical equivalences and alternative formulations increases the effective translation accuracy from $34.7\%$ to $77.8\%$.
    }
}
\vspace{0.15cm}

Overall, the results show that the pipeline progressively refines informal specifications into formally verifiable queries: the LLM effectively extracts functional requirements (\textbf{RQ1}), reliably filters unverifiable constraints (\textbf{RQ2}), and generates syntactically valid and semantically correct \uppaal queries (\textbf{RQ3}).

\section{Conclusion}
\label{sec:concl}

In this paper, we presented a pipeline that bridges informal natural language specifications and formal verification using \acp{llm}. The approach decomposes the process into three stages: (i) extraction of structured atomic requirements, (ii) classification of their formal verifiability with respect to the system model, and (iii) translation of verifiable requirements into UPPAAL \ac{smc} queries. Evaluation of three cyber-physical system scenarios shows that the pipeline effectively supports this transformation. The results indicate that \acp{llm} capture the core functional intent of informal specifications, reliably filter unverifiable requirements, and generate syntactically valid and semantically correct formal queries. Notably, semantic evaluation shows that strict string matching substantially underestimates formal query generation performance.

Future work will investigate intrinsic evaluation metrics to automatically assess intermediate outputs and enable refinement loops that improve each pipeline stage while mitigating error propagation. We also plan a broader empirical evaluation with additional \acp{llm}, more experimental repetitions, and a wider range of scenarios and models. Finally, we will explore fine-tuned models for requirement extraction and refine system prompt rules to address recurring \ac{llm} failure cases.

\section*{Acknowledgments}
\label{sec:acks}

The authors would like to thank Vincenzo Scotti for the provision of computational resources and expert guidance.

\printbibliography
\label{sec:bib}


\end{document}